\documentclass[aps,pre,superscriptaddress,amsmath,preprint,showpacs]{revtex4}
\usepackage{amsmath}
\usepackage[dvips]{graphicx}
\begin{document}
\title{Phase-space structures in quantum-plasma wave turbulence}

\author{F. Haas}
\altaffiliation{Universidade do Vale do Rio dos Sinos - UNISINOS, Av. Unisinos 950, 93022-000, S\~ao Leopoldo, RS, Brazil}
\affiliation{Institut f\"ur Theoretische Physik IV, Ruhr-Universit\"at Bochum\\ D-44780 Bochum, Germany}

\author{B. Eliasson}
\affiliation{Institut f\"ur Theoretische Physik IV, Ruhr-Universit\"at Bochum\\ D-44780 Bochum, Germany}
\affiliation{Department of Physics, Ume\aa \, University, SE-901 87, Ume{\aa}, Sweden}

\author{P. K. Shukla}
\altaffiliation{Department of Physics, Ume\aa \, University, SE-901 87, Ume{\aa}, Sweden}
\altaffiliation{GOLP / Instituto de Plasmas e Fus\~ao Nuclear, Instituto Superior T\'ecnico, Universidade T\'ecnica de Lisboa, 1049-001 Lisboa, Portugal}
\altaffiliation{SUPA, Department of Physics, University of Strathclyde, Glasgow, G40NG, UK}
\altaffiliation{School of Physics, University of Kwazulu-Natal, Durban 4000, South Africa.}
\affiliation{Institut f\"ur Theoretische Physik IV, Ruhr-Universit\"at Bochum\\ D-44780 Bochum, Germany}

\author{G. Manfredi}
\affiliation{Institut de Physique et Chimie des Mat\'eriaux de Strasbourg \\ BP43, F-67034 Strasbourg, France}

\begin{abstract}
\noindent The quasilinear theory of the Wigner-Poisson system in
one spatial dimension is examined. Conservation laws and
properties of the stationary solutions are determined. Quantum
effects are shown to manifest themselves in transient periodic
oscillations of the averaged Wigner function in velocity space.
The quantum quasilinear theory is checked against numerical
simulations of the bump-on-tail and the two-stream instabilities. The
predicted wavelength of the oscillations in velocity space agrees
well with the numerical results.
\end{abstract}
\maketitle

\section{Introduction}
Quantum plasmas have attracted a renewed attention in recent
years. The inclusion of quantum terms in the plasma fluid
equations -- such as quantum diffraction effects, modified
equations of state \cite{h1, m1}, and spin degrees of freedom
\cite{bm} -- leads to a variety of new physical phenomena. Recent
advances include linear and nonlinear quantum ion-acoustic waves
in a dense magnetized electron-positron-ion plasma \cite{Khan},
the formation of vortices in quantum plasmas \cite{Haque}, the
quantum Weibel and filamentation instabilities
\cite{h2}--\cite{Bret}, the structure of weak shocks in quantum
plasmas \cite{Bychkov}, the nonlinear theory of a quantum diode in
a dense quantum magnetoplasma \cite{b}, quantum ion-acoustic waves
in single-walled carbon nanotubes \cite{Weiw}, the many-electron
dynamics in nanometric devices such as quantum wells
\cite{MH1,MH2}, the parametric study of nonlinear electrostatic
waves in two-dimensional quantum dusty plasmas \cite{Ali},
stimulated scattering instabilities of electromagnetic waves in an
ultracold quantum plasma \cite{s}, and the propagation of waves
and instabilities in quantum plasmas with spin and magnetization
effects \cite{s1, s2}.

However, to date, only few works have investigated the important
question of quantum plasma turbulence. The most notable exception
is the paper by Shaikh and Shukla \cite{Shaikh}, where simulations
of the two- and three-dimensional coupled Schr\"odinger and
Poisson equations -- with parameters representative of the
next-generation laser-solid interaction experiments, as well as of
dense astrophysical objects -- were carried out. In that work, new
aspects of the dual cascade in two-dimensional electron plasma
wave turbulence at nanometric scales were identified.
Nevertheless, the quantum-plasma wave turbulence remains a largely
unexplored field of research. A reasonable strategy to attack
these problems would consist in extending well-known techniques
issued from the theory of classical plasma turbulence in order to
include quantum effects.

In this context, the simplest approach is given by the weak
turbulence kinetic equations first derived in Refs.
\cite{Vedenov1}--\cite{Drummond1}, the so-called quasilinear
theory. In quasilinear theory, the non-oscillating part of the
distribution function is flattened in the resonant region of
velocity space. It is interesting to carry over the basic
techniques of the classical quasilinear theory to the kinetic
models of quantum plasmas. The resulting quantum quasilinear
equations would be a useful tool for the study of quantum plasma
(weak) turbulence, and for quantum plasmas in general. In this
regard, it is natural to initially restrict the analysis to the
quasilinear relaxation of a one-dimensional quantum plasma in the
electrostatic approximation.

Some earlier works explored the similarities between the classical
plasma and the quantum-mechanical treatment of a radiation field
\cite{Pines}--\cite{Wei}. In those papers, one of the aims was to
obtain information on the classical plasma through a
quantum-mechanical language. For instance, the relaxation of an
instability can be viewed as the spontaneous emission of ``quanta"
of a radiation field described by some quasilinear-type equations.
However, the application of the quasilinear method to a truly
quantum plasma governed by the Wigner-Poisson system (i.e. the
quantum analogue of the Vlasov-Poisson system) seems to be
restricted to the work of Vedenov \cite{Vedenov}. Surprisingly,
there has been no systematic analysis of the consequences of the
quasilinear theory in the Wigner-Poisson case. The present work is
a first attempt in this direction.

This manuscript is organized in the following fashion. In Sect. 2,
the quantum quasilinear equations are derived from the
Wigner-Poisson system. In comparison with the classical
quasilinear equations, the quantum model exhibits a
finite-difference structure. The basic properties of the quantum
quasilinear theory are discussed in Sect. 3, where we derive some
conservation laws, as well as an appropriate H-theorem for the
quasilinear equations. The existence of an entropy-like quantity
is used to prove that the averaged Wigner function relaxes to a
plateau, just like in the classical case. However, the distinctive
feature of the quantum quasilinear equations is the existence of a
transient periodic structure in velocity space, as shown in Sect.
4. In Sect. 5, the theory is checked against numerical simulations
of the bump-on-tail and two-stream instabilities, with good
agreement with the predictions. Conclusions are drawn in Sect. 6.

\section{Quantum quasilinear equations}
The quasilinear equations for the Wigner-Poisson system were
derived long ago by Vedenov \cite{Vedenov}, without fully
exploring their consequences. For completeness, the derivation
procedure is reproduced here. The Wigner equation reads
\begin{equation}
\label{eqq1} \frac{\partial{f}}{\partial{t}} + v
\frac{\partial{f}}{\partial{x}} = \int dv' K(v-v',x,t)f(x,v',t) \,,
\end{equation}
where
\begin{equation}
K(v-v',x,t) = -\frac{iem}{2\pi\hbar^2} \int d\lambda \,
e^{im(v-v')\lambda/\hbar}
\left[\phi(x+\lambda/2,t)-\phi(x-\lambda/2,t)\right] \,.
\label{wig_acc}
\end{equation}
Here, $f(x,v,t)$ is the Wigner pseudo-distribution in one spatial
dimension, with position $x$, velocity $v$, and time $t$. Also,
$\hbar = h/2\pi$ is the scaled Planck's constant, $e$ is the absolute
value of the electron charge, and $m$ is the electron mass.
The electrostatic potential $\phi(x,t)$ satisfies the Poisson
equation,
\begin{equation}
\label{po} \frac{\partial^{2}\phi}{\partial x^2} =
\frac{e}{\varepsilon_{0}}\left(\int dv f - n_{0}\right) \, ,
\end{equation}
where $\varepsilon_0$ is the vacuum permittivity and
$n_0$ a fixed, neutralizing ionic background. Periodic boundary
conditions are assumed, with periodicity length $L$. Accordingly,
for any quantity $A = A(x,v,t)$, the spatial average is
\begin{eqnarray}
\langle A(x,v,t) \rangle = \frac{1}{L}\int_{-L/2}^{L/2}dx A(x,v,t)
\,.
\end{eqnarray}
In particular, it is useful to define $F(v,t) = \langle f(x,v,t)
\rangle$ and to restrict to $ \langle \phi(x,t) \rangle = 0$.

The quasilinear theory proposes a perturbation solution of the form
\begin{equation}
f = F(v,t) + f_{1}(x,v,t) \,, \quad \phi = \phi_{1}(x,t) \,,
\end{equation}
for small $f_1$ and $\phi_1$.  After averaging the Wigner
equation, taking into account that $\langle\phi_{1}(x,t)\rangle =
0$, we obtain from (\ref{eqq1})
\begin{equation}
\label{eq1} \frac{\partial F}{\partial t} =
\frac{iem}{2\pi\hbar^2}\int d\lambda dv' e^{im(v-v')\lambda/\hbar}
\left\langle\left[\phi_{1}\Bigl(x - \frac{\lambda}{2},t\Bigr)
- \phi_{1}\Bigl(x + \frac{\lambda}{2},t\Bigr)\right] f_{1}(x,v',t)\right\rangle  \,.
\end{equation}
In the quasilinear theory, the right-hand side of Eq. (\ref{eq1}) is
evaluated using the results of the linear theory. This implies, in
particular, that the mode coupling effects are not taken into account.
Notice that $F$ changes slowly, since $\partial F/\partial t$ is a
second order quantity. Thus, for too strong damping or
instability, the quasilinear theory is no longer valid. Also, the trapping
effect can be included only in the framework of a fully nonlinear
theory. Generally speaking, the conditions of validity of the
quasilinear theory are still the subject of hot debates at the
classical level \cite{laval}.

We now introduce the spatial Fourier transforms
\begin{eqnarray}
\hat{f}_{1k}(v,t) &=& \frac{1}{\sqrt{2\pi}}\int_{-L/2}^{L/2} dx \,e^{-ikx} f_{1}(x,v,t)  \,, \\
\hat{\phi}_{1k}(t) &=& \frac{1}{\sqrt{2\pi}} \int_{-L/2}^{L/2} dx\,e^{-ikx} \,\phi_{1}(x,t) \,,
\end{eqnarray}
with the corresponding inverse transforms
\begin{eqnarray}
f_{1}(x,v,t) &=& \frac{\sqrt{2\pi}}{L}\sum_{k} e^{ikx} \hat{f}_{1k}(v,t)  \,, \\
\phi_{1}(x,t) &=& \frac{\sqrt{2\pi}}{L}\sum_{k} e^{ikx} \hat{\phi}_{1k}(t) \,,
\end{eqnarray}
where $k = 2\pi n/L$, $n = 0, \pm 1, \pm 2,...$
Linearizing the Wigner equation, Fourier transforming it in space and Laplace transforming it in time, we obtain
\begin{equation}
\label{x3}
\hat{f}_{1k}(v,t) = \frac{em \hat{\phi}_{1k}(t)}{2\pi\hbar^{2}(\omega_k - kv)} \int
d\lambda\, dv'\, e^{im(v-v')\lambda/\hbar} (e^{ik\lambda/2} - e^{-ik\lambda/2})\,F(v',t) \,.
\end{equation}
In Eq. (\ref{x3}), since we are interested only in the long-lived collective oscillations,
the initial perturbation $f_{1}(x,v,0)$ was neglected. Finally,
$\omega_k$ stands for the allowable frequency modes, obtained from
the the well-known \cite{Drummond} quantum dispersion relation
\begin{equation}
\label{disp} D(k,\omega) = 1 - \frac{\omega_{p}^2}{n_{0}} \int_{{\cal L}}
dv\, \frac{F(v)}{(\omega_k - kv)^2 - \hbar^{2}k^{4}/4m^2} = 0 \,,
\end{equation}
which is assumed to be adiabatically valid, where ${\cal L}$ denotes the Landau contour, and $\omega_{p} =
(n_{0}e^{2}/m\varepsilon_{0})^{1/2}$ is the electron plasma frequency; for brevity, we omit to write out
the second argument $t$ of $F$ from here and onward.

Since $f_1$ and  $\phi_1$ are real, we have the parity properties
\begin{equation}
\label{pp}
\hat{f}_{1,-k}(v,t) = \hat{f}_{1k}^{*}(v,t) \,, \quad
\hat{\phi}_{1,-k}(t) = \hat{\phi}_{1k}^{*}(t) \,, \quad
\omega_{-k} = - \omega^{*}_k \,,
\end{equation}
which will be used in the remainder of this paper. By
defining
\begin{equation}
\omega_k = \Omega_k + i\gamma_k \,,
\end{equation}
where $\Omega_k$ and $\gamma_k$ are real,  the following
properties also hold
\begin{equation}
\Omega_{-k} = - \Omega_k \,, \quad \gamma_{-k} = \gamma_k \,.
\label{pp2}
\end{equation}

By using Eqs. (\ref{x3}) and (\ref{pp}) into (\ref{eq1}), and carrying out
straightforward calculations, we obtain
\begin{equation}
\label{x4} \frac{\partial F}{\partial t} = \frac{2\pi ie^2}{L^2\hbar^2}\sum_{k\neq 0} |\hat{\phi}_{1k}(t)|^2 \,\left[\frac{F(v+\hbar k/m)-F(v)}{\omega_k-kv-\hbar k^{2}/2m}+\frac{F(v-\hbar k/m)-F(v)}{\omega_k-kv+\hbar k^{2}/2m}\right] \,.
\end{equation}
Equation (\ref{x4}) can be put in a convenient form by using
the parity properties (\ref{pp})--(\ref{pp2}) and
that $|\gamma_{k}|$ is small so that
$\gamma_{k}/[(\Omega_{k}-kv\pm\hbar k^2/2m)^2 + \gamma_{k}^2]
\simeq \pi\delta(\Omega_k - kv\pm\hbar k^2/2m)$, where $\delta$
is Dirac's delta function. Hence we express Eq. (\ref{x4}) as
\begin{eqnarray}
\frac{\partial F}{\partial t} &=& \frac{4\pi^2 m\omega_{p}^2}{n_0
L\hbar^2}\sum_{k>0} \frac{\hat{\varepsilon}_{k}(t)}{k^2}
\Bigl\{\Bigl[F\Bigl(v+\frac{\hbar k}{m}\Bigr) - F(v)\Bigr]
\delta\Bigl(\Omega_k-kv-\frac{\hbar k^{2}}{2m}\Bigr) \nonumber \\ \label{x5}
&+& \Bigl[F\Bigl(v-\frac{\hbar k}{m}\Bigr)-F(v)\Bigr]\delta\Bigl(\Omega_k-kv+\frac{\hbar
k^{2}}{2m}\Bigr)\Bigr\} \,,
\end{eqnarray}
in which the growth rate does not appear explicitly, and where we
have defined the spectral density of the electrostatic field
fluctuations as
\begin{equation}
\hat{\varepsilon}_{k}(t) =
\frac{\varepsilon_{0}}{L} k^{2}|\hat{\phi}_{1k}(t)|^{2}  \,.
\end{equation}
The presence of the delta functions in Eq. (\ref{x5}) emphasizes
the fact that only particles satisfying the resonance condition
$\Omega_k - kv \pm \hbar k^2/2m = 0$ are taken into account,
whereas $F(v,t)$ remains unchanged in the non-resonant region of
velocity space.

The time variation of the spectral density is given in the same way as in the
Vlasov-Poisson case, i.e.
\begin{equation}
\label{eq4} \frac{\partial\hat{\varepsilon}_k}{\partial t} =
2\gamma_k \hat{\varepsilon}_k \,.
\end{equation}
Equations (\ref{x5}) and (\ref{eq4}) are the quasilinear
equations for the Wigner-Poisson system. In the formal classical
limit $\hbar \rightarrow 0$, they reduce to the well-known
quasilinear equations for the Vlasov-Poisson system.

\section{Properties of the quantum quasilinear equations}
Taking velocity moments of Eq. (\ref{x5}), several conservation
laws can be easily derived. For instance, we obtain the
conservation of the number of particles,
\begin{equation}
\frac{d}{dt}\int\,dv\,F = 0 \,,
\end{equation}
and of the linear momentum
\begin{equation}
\frac{d}{dt}\int\,dv\,mv\, F =  0\,,
\end{equation}
where the dispersion relation (\ref{disp}) and the parity
properties of the spectral density have been used.  In addition,
the total energy is also invariant
\begin{equation}
\label{ene}
\frac{d}{dt}\left(\int\,dv\,\frac{mv^2}{2}\,F +\frac{2\pi}{L}\sum_{k}
\hat{\varepsilon}_{k}\right) = 0 \,.
\end{equation}
Intermediate steps to derive Eq. (\ref{ene}) require the use of
the parity properties as well as the second quasilinear equation (\ref{eq4}). In addition, a useful approximation is to
consider $\Omega_k \simeq \omega_p$ for $k>0$, jointly with
\begin{equation}
\label{gam} \gamma_k \simeq \frac{\pi m\omega_{p}^3}{2n_0 \hbar
k^3} \left[F\left(\frac{\omega_{p}}{k} + \frac{\hbar k}{2m}\right)
- F\left(\frac{\omega_{p}}{k} - \frac{\hbar k}{2m}\right)\right] \,.
\end{equation}

In order to gain insight into the asymptotic behavior of $F(v,t)$,
it is interesting to look for an entropy-like quantity. Following
Ref. \cite{manfredi}, we consider the quantity $\int F^2 dv$,
which, from Eq. (\ref{x5}), can be proved to obey the equation
\begin{equation}
\label{ent} \frac{d}{dt}\int dv F^2 = - \frac{8\pi^2
m\omega_{p}^2}{n_0 L\hbar^2} \sum_{k>0}
\frac{\hat{\varepsilon}_{k}(t)}{k^3}
\left[F\left(\frac{\Omega_{k}}{k} + \frac{\hbar k}{2m}\right) -
F\left(\frac{\Omega_{k}}{k} - \frac{\hbar k}{2m}\right)\right]^2
\leq 0 \,,
\end{equation}
where the last inequality follows since all terms in the right-hand
side are non-positive. This result constitutes a sort of H-theorem
for the averaged distribution function.

The time derivative of the non-negative quantity in the left-hand
side of Eq. (\ref{ent}) is always non-positive, so that
asymptotically we have $\frac{d}{dt}\int dv F^2 \to 0$, and
\begin{equation}
\label{a} F\left(\frac{\Omega_{k}}{k} + \frac{\hbar k}{2m}\right)
- F\left(\frac{\Omega_{k}}{k} - \frac{\hbar k}{2m}\right) = 0 \,,
\end{equation}
for all wavenumbers where the spectral density is non-zero.
Equation (\ref{a}) also resembles the basic equation obtained when
applying the Nyquist method to the stability analysis of the
Wigner-Poisson system \cite{nyquist}.

Equation  (\ref{a}) is a finite-difference-like version of the
classical plateau condition [$F'(v) = 0$] in the region where the
spectral density is not zero. The finite-difference structure of
Eq. (\ref{a}) favors the appearance of oscillations in velocity
space, which are not present in the classical case. These
oscillations are analyzed in detail in the next section.

\section{Transient quantum oscillations in velocity space}
In the derivation at the end of Sect. 3, it was implicitly assumed
that the spectral density is not zero in a broad region of
momentum space. In contrast, let us see what happens in the
idealized situation where the spectral density
$\hat{\varepsilon}_k$ is strongly peaked at a single mode $K$.
Denoting the associated spectral density by $\hat{\varepsilon}(t)$
and using the growth rate (\ref{gam}), the quantum quasilinear
equations read
\begin{eqnarray}
\label{x15} \frac{\partial F}{\partial t} &=&
\frac{mL\omega_{p}^2}{n_0 \hbar^2} \, \hat{\varepsilon}(t)
\Bigl\{\Bigl[F\Bigl(v+\frac{\hbar K}{m}\Bigr) - F(v)\Bigr]
\delta\Bigl(\omega_p-Kv-\frac{\hbar K^{2}}{2m}\Bigr)  \\
 &+& \Bigl[F\Bigl(v-\frac{\hbar K}{m}\Bigr)-F(v)\Bigr]
 \delta\Bigl(\omega_p-Kv+\frac{\hbar K^{2}}{2m}\Bigr)\Bigr\} \,, \nonumber
\end{eqnarray}
and
\begin{equation}
\label{den} \frac{d\hat{\varepsilon}(t)}{dt} =  \frac{\pi
m\omega_{p}^3}{n_0 \hbar K^3} \,\hat{\varepsilon}(t)
\left[F\left(\frac{\omega_{p}}{K} + \frac{\hbar K}{2m}\right) -
F\left(\frac{\omega_{p}}{K} - \frac{\hbar K}{2m}\right)\right] \,,
\end{equation}
where, for simplicity, the approximation $\Omega_K = \omega_p$ was
also adopted.

A particular class of stationary solutions of Eqs.
(\ref{x15})-(\ref{den}) is given by any function $F(v)$ that is a
periodic in velocity space, with period $\hbar K/m$:
\begin{equation}
\label{denn2} F\left(v+\hbar K/2m\right) - F\left(v-\hbar
K/2m\right) = 0.
\end{equation}
Assuming $F(v) \sim \exp(i\alpha v)$ in Eq. (\ref{denn2}), with
$\alpha$ to be determined, we easily obtain the characteristic
equation $\sin(\alpha \hbar K/2m) = 0$. Hence, the general (exact)
equilibrium solution is the linear combination
\begin{equation}
\label{exa} F(v) = a_0 + \sum_{n=1}^{\infty} a_{n}
\cos\left(\frac{2\pi n v}{\lambda_v}\right) + \sum_{n=1}^{\infty}
b_{n} \sin\left(\frac{2\pi n v}{\lambda_v}\right)\,,
\end{equation}
where $a_{n}, b_{n}$ are arbitrary real constants. Notice the
singular character of the quantum oscillations, whose
``wavelength'' of the fundamental mode ($n = 1$) in velocity
space, $\lambda_v = \hbar K/m$, tends to zero as $\hbar
\rightarrow 0$. The solution given by Eq. (\ref{exa}) represents
periodic oscillations in velocity space. However, this is
necessarily a transient solution that cannot be sustained for long
times. Indeed, strictly speaking, Eq. (\ref{exa}) applies only at
the resonance, which is a set of measure zero on the velocity
axis. The generation of harmonics with wavenumbers $k\neq K$
(which is forbidden in the quasilinear theory) would ultimately
lead to a broad energy spectrum.

When many wavenumbers are present, the simple periodic solution
given in Eq. (\ref{denn2}) does not hold anymore, because
different values of $k$ induce different velocity-space
wavelengths $\lambda_v$. Only the solution $F(v) = \rm const.$
holds independently of $k$. Therefore, we expect that a plateau
will eventually appear on a finite region in velocity space.
Nevertheless, Eq. (\ref{exa}) provides an estimate for the
characteristic oscillation length  of the averaged Wigner function
in velocity space, near resonance.

The above arguments also suggest that monochromatic waves are the
best candidates for displaying such quantum oscillations. In
contrast, if the energy spectrum is broad from the very start, the
formation of a plateau is likely to be favored over the appearance
of periodic oscillations. In the following section, these
predictions will be compared to numerical simulations of the
Wigner-Poisson system.

\section{Numerical simulations}
We have performed numerical simulations of the Vlasov and Wigner
equations using a phase-space code based on a splitting method
\cite{Suh}. In the Wigner equation, the acceleration term
(\ref{wig_acc}) is a convolution product in velocity space and
is therefore calculated numerically by Fourier transforming it in velocity space.
\begin{figure}
\centering
\includegraphics[width=6cm]{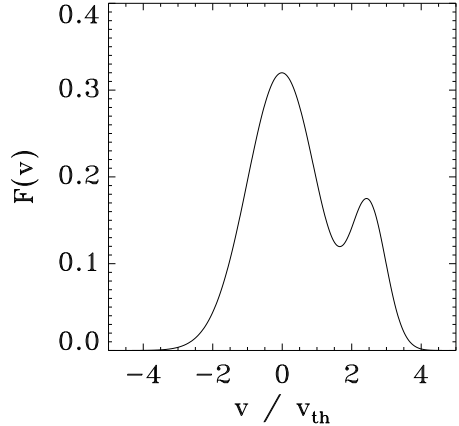}
\caption{Initial velocity distribution used in the simulations of
the bump-on-tail instability, both for the Vlasov and Wigner cases.}
\end{figure}
\begin{figure}
\includegraphics[width=4.5cm]{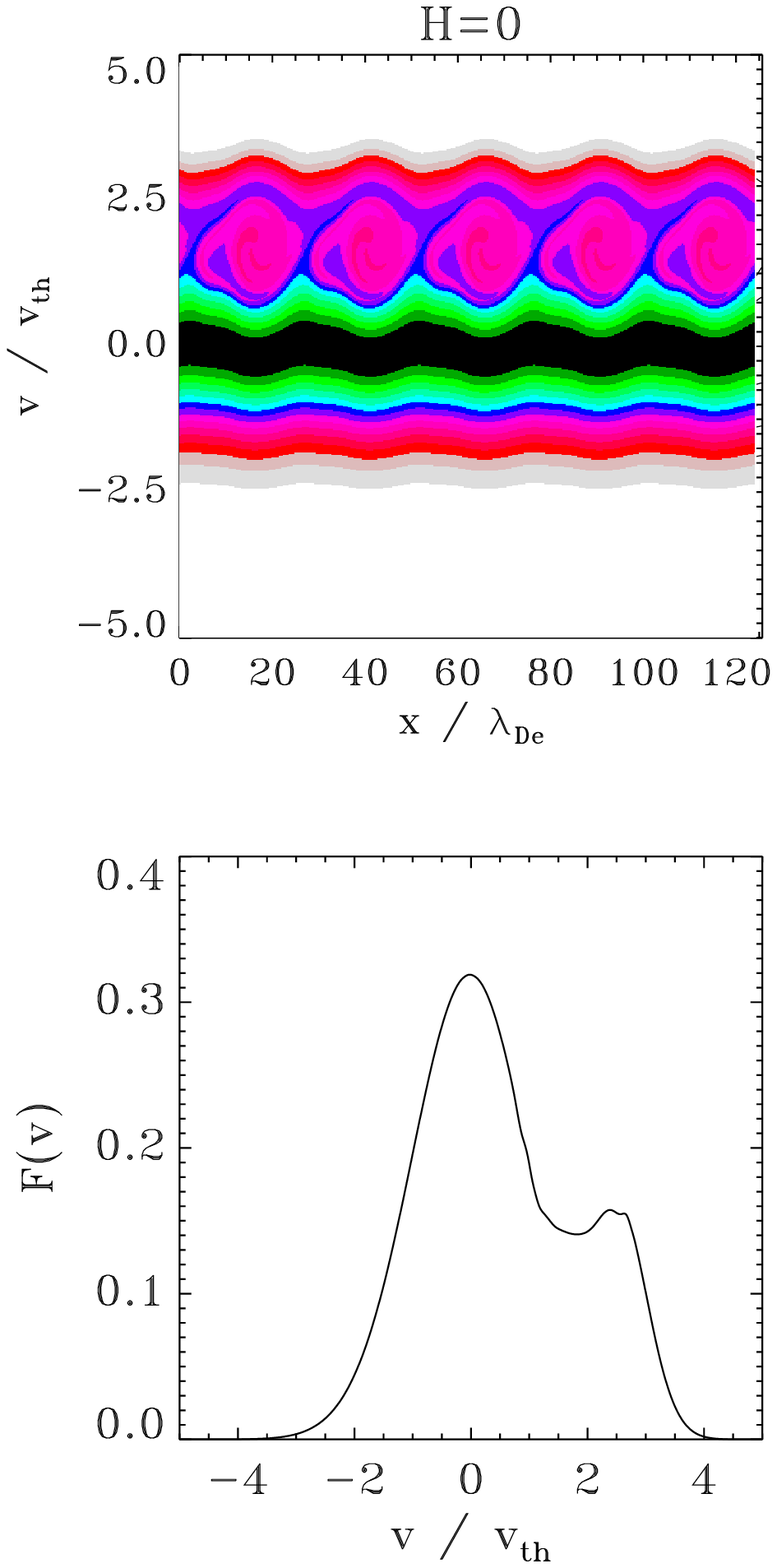}
\includegraphics[width=4.5cm]{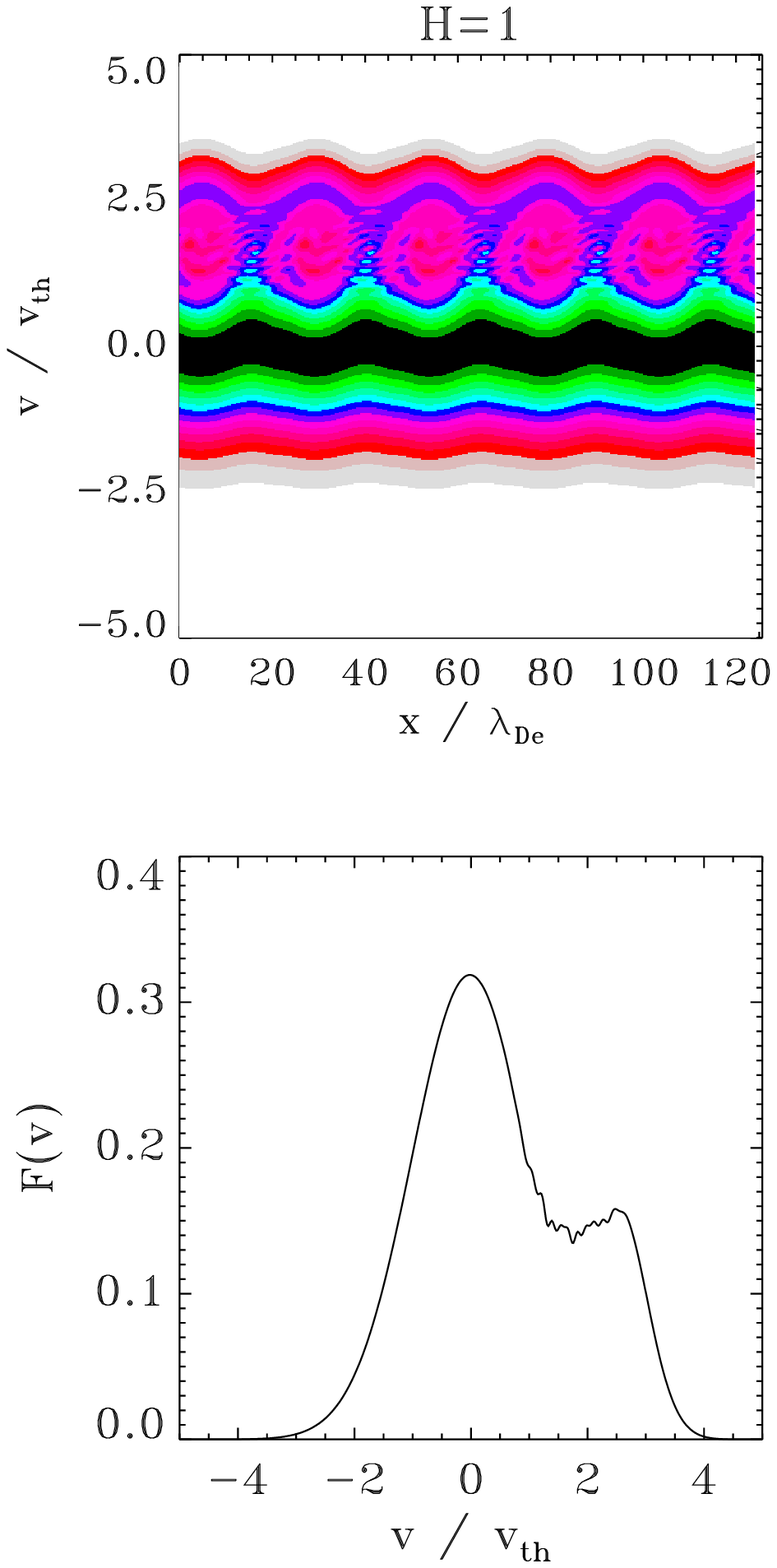}
\includegraphics[width=4.5cm]{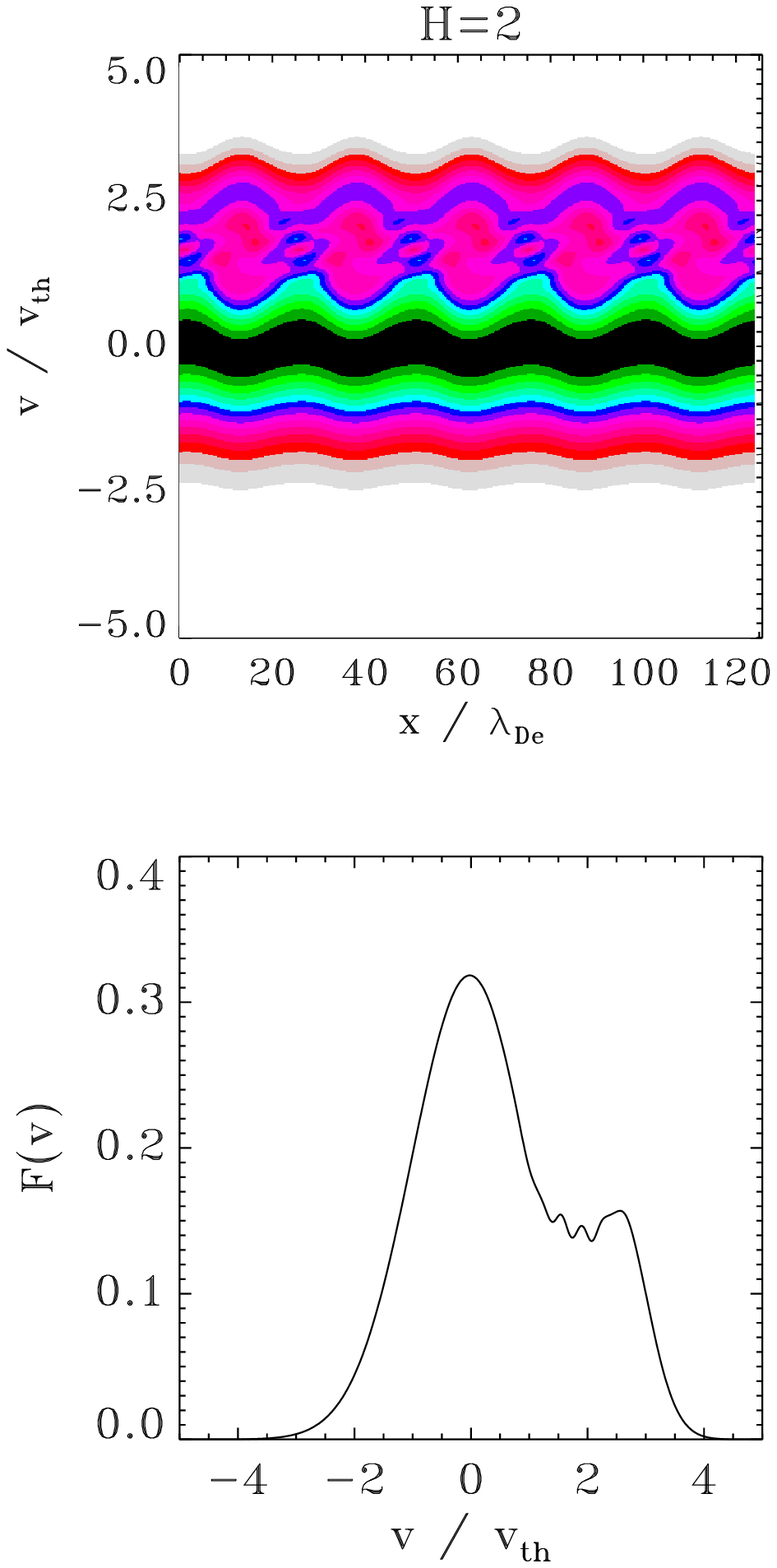}
\caption{Simulations of the Wigner-Poisson and Vlasov-Poisson
systems, at time $\omega_p t =200$, for $H=0$ (Vlasov, left
frame), $H=1$ (middle frame), and $H=2$ (right frame). Initially
monochromatic spectrum. Top panels: electron distribution function
$f(x,v)$ in phase space. Bottom panels: spatially averaged
electron distribution function $F(v)$ in velocity space.}
\end{figure}

To study the differences in the nonlinear evolution of the Wigner
and Vlasov equations, we have simulated the well-known
bump-on-tail instability, whereby a high-velocity beam is used to
destabilize a Maxwellian equilibrium. We use the initial condition
$f=(1+\delta)(n_0/\sqrt{2\pi} v_{th}) [0.8 \exp(-v^2/2 v_{th}^2) +
0.4 \exp(-2(v-2.5v_{th})^2/v_{th}^2)]$, where $\delta$ represents random
fluctuations of order $10^{-5}$ that help seed the instability
(see Fig. 1). Here $v_{th}=\sqrt{k_B T_e/m}$ is the electron
thermal speed. We use periodic boundary conditions with spatial
period $L=40\pi\lambda_{De}$, where
$\lambda_{De}=v_{th}/\omega_{p}$ is the Debye length. Three
simulations were performed, with different values of the
normalized Planck constant, defined as
$H=\hbar\omega_p/mv_{th}^2$: $H=0$ (Vlasov), $H=1$, and $H=2$.

In order to highlight the transient oscillations in velocity
space, we first perturb the above equilibrium with a monochromatic
wave having $\lambda_{De}k= 0.25$ (i.e., a wavelength of
$8\pi\lambda_{De}$). Figures 2 shows the results from simulations
of the Wigner-Poisson and Vlasov-Poisson systems. In both
simulations, due to the bump-on-tail instability, electrostatic
waves develop nonlinearly and create periodic trapped-particle
islands (electron holes) with the wavenumber $k= 0.25$.
The theory described in the previous sections predicts the
formation of velocity-space oscillations in the Wigner evolution,
which should be absent in the classical (Vlasov) simulations. This
is the case in the results presented in Fig. 2, where the
oscillations are clearly visible.

In order to estimate their wavelength, a zoom on the spatially
averaged electron distribution function $F(v)$ is shown in Fig. 3.
According to the quasilinear theory, the velocity wavelength should be
equal to $\lambda_v = \hbar k/m$ for the fundamental
mode with $n=1$ [see Eq. (\ref{exa})]. In our units, this yields
$\lambda_v/v_{th} = H k\lambda_{De} $, which is equal to 0.25 for
$H=1$ and to 0.5 for $H=2$. The wavelengths observed in the
simulations are slightly smaller: $\lambda_v/v_{th} \simeq 0.17$
and 0.35 for $H=1$ and $H=2$, respectively. However: (i) the
oscillations are absent in the Vlasov case, as expected, (ii) the
order of magnitude of the wavelength is correct, and (iii) the
wavelength is proportional to $H$, in accordance with the quasilinear
theory. The slight discrepancy in the observed value of
$\lambda_v$ may have at least two origins. First, spatial
wavenumbers different from $0.25$ can be excited
due to the nonlinear mode coupling. Second, other modes with $n>1$
[see Eq. (\ref{exa})] can affect the velocity-space wavelength.
\begin{figure}
\centering
\includegraphics[width=6cm]{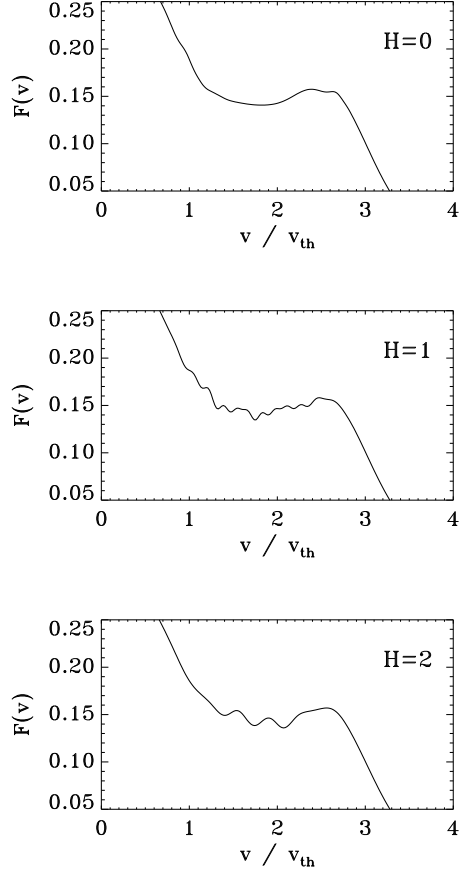}
\caption{Zoom on the spatially averaged electron distribution
function $F(v)$ in velocity space for the same cases as in Fig.
2.}
\end{figure}

When the initial excitation is broad-band (i.e., wavenumbers $0.05
\le k \le 0.5$ are excited), the electron holes start
merging together at later times due to the sideband instability
\cite{Kruer69,Albrecht07} (see Fig. 4). At this stage, mode
coupling becomes important and quasilinear theory is not capable
of describing these effects. As the system evolves toward larger
spatial wavelength, the evolution becomes progressively more
classical, with the appearance of a plateau in the resonant
region. Nevertheless, at $\omega_p t =500$ the Wigner solution
still displays some oscillatory behavior in velocity space, which
is absent in the Vlasov evolution.
\begin{figure}
\centering
\includegraphics[width=5cm]{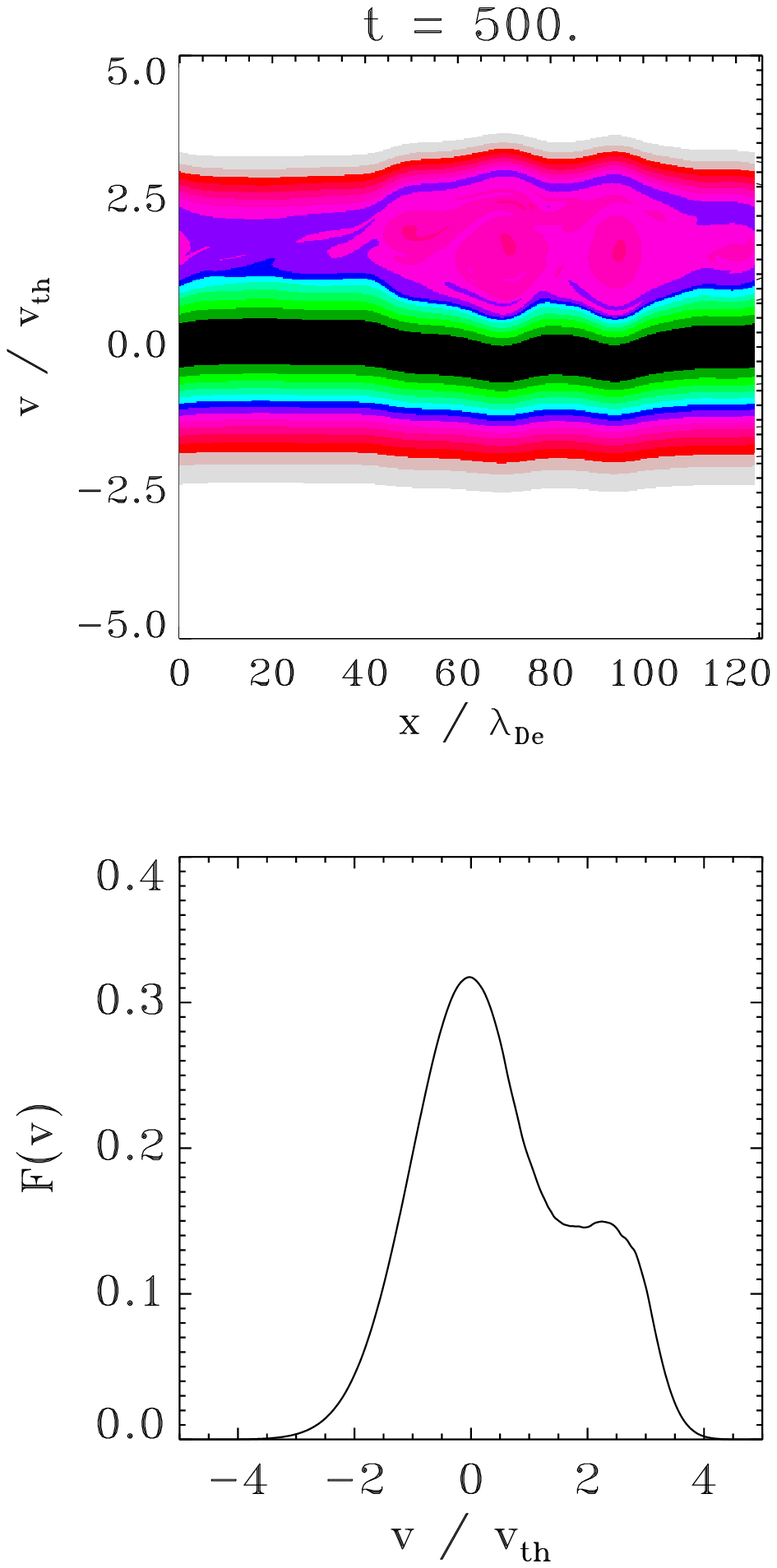}
\includegraphics[width=5cm]{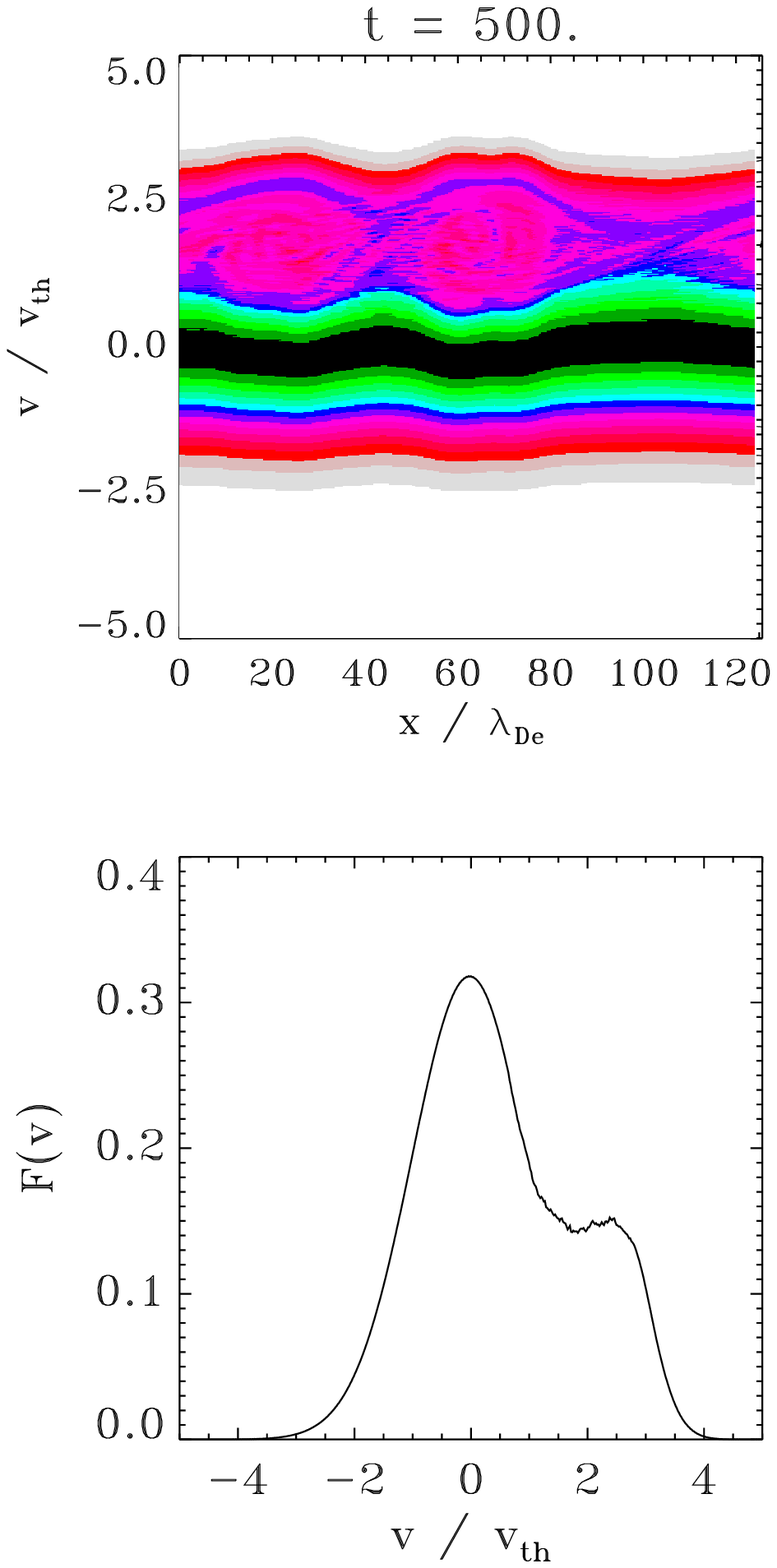}
\caption{Simulations of the Wigner-Poisson and Vlasov-Poisson
systems, for $\omega_p t =500$, for $H=0$ (Vlasov, left panel)
and $H=1$ (Wigner, right panel). Initially broad wavenumber
spectrum. Top panels: electron distribution function $f(x,v)$ in
phase space. Bottom panels: spatially averaged electron
distribution function $F(v)$ in velocity space.}
\end{figure}

Another set of simulations were performed for the case of a
two-stream instability. The initial distribution function is composed
of two Maxwellians with thermal speed $v_{th}$, each centered
at $v=\pm 2v_{th}$ (see Fig. 5). Only the fundamental mode of the
system, with the wavenumber $K=0.2\lambda_{De}^{-1}$, is excited, and
it grows exponentially due to instability. Three simulations
were performed, with different values of the normalized Planck
constant: $H=0$ (Vlasov), $H=1$, and $H=2$. The expected
oscillation wavelength in velocity space is $\lambda_v=\hbar K/m$.
For the three simulated cases, it should take the values
$\lambda_v=0$, $\lambda_v=0.2v_{th}$, and $\lambda_v=0.4v_{th}$. A
zoom of the averaged Wigner function $F(v,t)$ around the velocity
$v=0$ is plotted in Fig. 6 for the three cases, at time $\omega_p
t=110$. The velocity-space oscillations are absent from the Vlasov
simulation. In the Wigner cases, their wavelength is rather close
to the theoretical value; in particular, it appears to grow with
the scaled Planck constant, as expected from the theory. As in the
bump-on-tail case, the oscillations tend to disappear over longer
times.
\begin{figure}
\centering
\includegraphics[width=6cm]{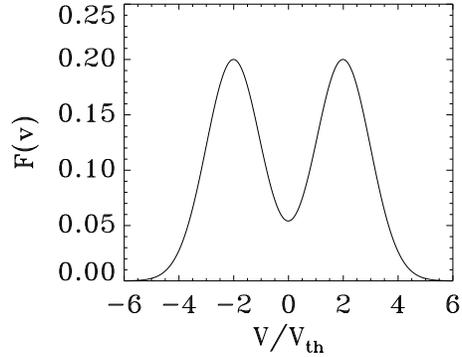}
\caption{Initial velocity distribution used in the simulations of
the two-stream instability, both for the Vlasov and Wigner cases.}
\end{figure}
\begin{figure}
\centering
\includegraphics[width=8cm]{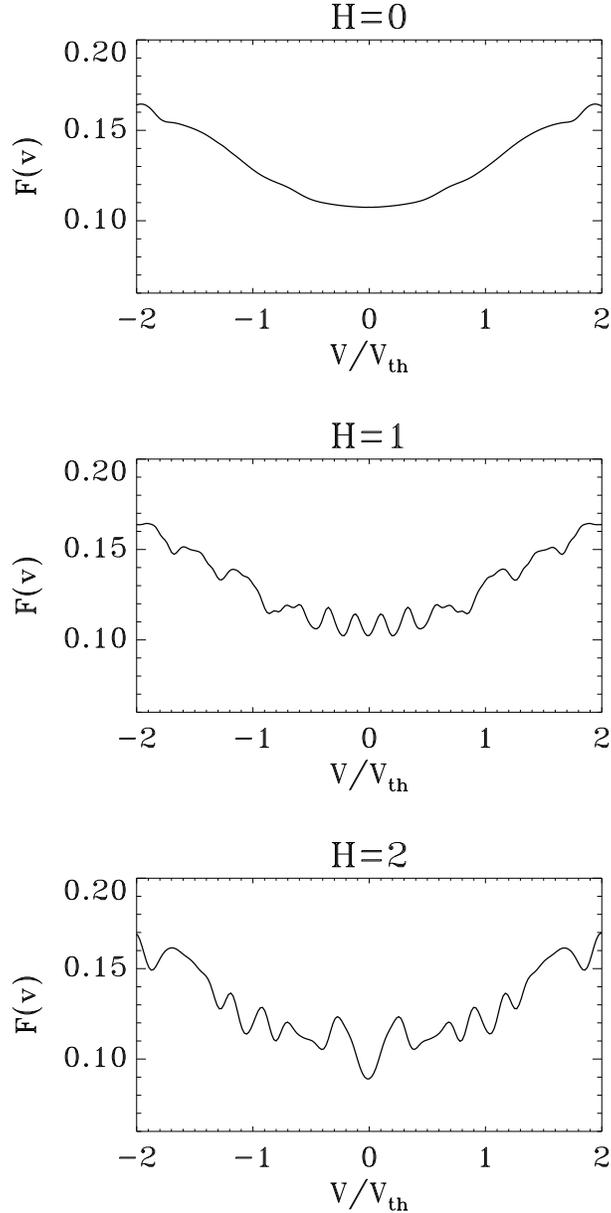}
\caption{Zoom of the averaged Wigner distribution $F(v)$ for the
two-stream instability at time $\omega_p t=110$. Top frame: $H=0$
(Vlasov); middle panel: $H=1$; bottom panel: $H=2$.}
\end{figure}

\section{Conclusion}
In this paper, the quasilinear theory for the Wigner-Poisson system
was revisited. Conservation laws and the asymptotic solution were
established. Distinctive quantum effects appear in the form of a
transient oscillatory behavior in velocity space. Such quantum
effects are favored when the energy spectrum is restricted to few
modes. For longer times, the plasma tends to become classical --
due to the spatial harmonic generation and the mode couplings -- at least
as far as the averaged Wigner function $F(v,t)$ is concerned.
Thus, just as in the classical case, $F(v,t)$ evolves
asymptotically toward a plateau in the resonant region of
velocity space. It would be interesting to investigate whether
monochromaticity enhances quantum effects in plasmas in general,
which would represent a useful property for an experimental
validation of the quantum plasma models.

\vskip .5cm {\bf Acknowledgments} \vskip .5cm

This work was partially supported by the Alexander von Humboldt
Foundation and by the Swedish Research Council.

\end{document}